\documentclass[preprint,showpacs,preprintnumbers,amsmath,amssymb]{revtex4}

\usepackage{graphicx}
\usepackage{dcolumn}
\usepackage{bm}

\begin{document}

\preprint{Otani2003}

\title{First-principles study of electron transport through C$_{20}$ cages}

\author{Megumi Otani}
\affiliation{Department of Precision Science and Technology, Osaka University, Suita, Osaka 565-0871, Japan}

\author{Tomoya Ono}
\affiliation{Research Center for Ultra-Precision Science and Technology, Osaka University, Suita, Osaka 565-0871, Japan}

\author{Kikuji Hirose}
\affiliation{Department of Precision Science and Technology, Osaka University, Suita, Osaka 565-0871, Japan }

\date{\today}


\begin{abstract}
Electron transport properties of C$_{20}$ molecules suspended between gold electrodes are investigated using first-principles calculations. Our study reveals that the conductances are quite sensitive to the number of C$_{20}$ molecules between electrodes: the conductances of C$_{20}$ monomers are near 1 G$_{0}$, while those of dimers are markedly smaller, since incident electrons easily pass the C$_{20}$ molecules and are predominantly scattered at the C$_{20}$-C$_{20}$ junctions.
Moreover, we find both channel currents locally circulating the outermost carbon atoms.

\end{abstract}

\maketitle


As new techniques for the atomic-scale manipulation and modification of materials progress, electron transport properties of nanostructures have attracted considerable interest \cite{Datta}.
To date, special attention has been focused on atomic and molecular systems, such as organic molecules and fullerenes, because they are expected to be the ultimate size limit of functional devices and show unique behaviors different from those of macroscopic systems \cite{Molecular}.
Joachim {\it et al.} \cite{Joachim} indicated experimentally that a deformed C$_{60}$ molecule has the possibility of being a nanoscale electrical amplifier.
Recently, a C$_{20}$ cage solely with pentagons has been produced from C$_{20}$H$_{20}$ \cite{Prinzbach}, which is the smallest fullerene that is one of the candidates of much minuter electronic devices.
While these prospects of fullerene-based devices are exciting, there remains much to be learned about their electronic characteristics.

On the theoretical side, first-principles calculations have been employed to provide insight into geometric and electronic properties of nanostructures.
Concerning C$_{20}$ cage systems, Miyamoto {\it et al.} \cite{miyamoto} studied C$_{20}$ one-dimensional infinite chains to indicate that some stable chain structures can be formed and one of them possesses semiconductivity.
Later, Roland {\it et al.} \cite{Roland} presented electron-conduction properties of one-dimensional short chains made of several C$_{20}$ molecules suspended between Al or Au leads, and they summarized that the conductances do not depend on the number of molecules contained in the chains.
Although these studies give us certain knowledge about fullerene-based devices, a transparent view on the quantized electron transport, which can contribute to designing future devices, has not been provided yet, and further examinations of the conduction properties and the current flowing through the devices are indispensable.

In this Rapid Communication, we present elaborate first-principles calculations for electron transport properties, in particular, the conduction channels and their current distributions of C$_{20}$ molecules suspended between two semi-infinite Au electrodes.
To the best of our knowledge, this is the first report focusing on how incident electrons can flow and reflect within the system of C$_{20}$ molecules intervening between the electrodes.
Our main findings are as follows: (1) contrary to the conclusion reported by Roland {\it et al.} \cite{Roland}, the number of C$_{20}$ molecules between the electrodes largely affects the electron transport behavior. 
The conductances of C$_{20}$ {\it monomers} are around 1 G$_{0}$(=2$e^{2}/h$), and on the other hand, those of {\it dimers} are markedly smaller.
(2) Electron currents are found to pass along the C-C bonds, and not to directly cross through to inside the C$_{20}$ cages.
Backscattering of incident electrons occurs at both the C$_{20}$-electrode interfaces and the C$_{20}$-C$_{20}$ junctions of the dimers, while reflection seldom takes place within the C$_{20}$ molecules.
(3) Channel current distributions exhibit local loops around the outermost carbon atoms.


Our first-principles molecular-dynamics simulation is based on the real-space finite-difference method \cite{rsfd}, which enables us to determine the self-consistent electronic ground state and the optimized atomic geometry with a high degree of accuracy, by making use of the timesaving double-grid technique \cite{tsdg} and the direct minimization of the energy functional \cite{dmef}.
The norm-conserving pseudopotentials \cite{ncps} of Troullier and Martins \cite{tmpp} are adopted and exchange-correlation effects are treated by the local density approximation \cite{lda} of the density functional theory.
The electron transmission at the Fermi level is calculated by the Landauer formula in cooperation with the overbridging boundary-matching method \cite{fujimoto}.
Figure \ref{fig:models} shows the calculation models employed here.
For C$_{20}$ monomers sandwiched between electrodes, we examine the two models, a double-bonded one shown in Fig.\ref{fig:models}(a) and a single-bonded one shown in Fig.\ref{fig:models}(b).
In the cases of dimers, we adopt the double-bonded and single-bonded dimers indicated in Figs.\ref{fig:models}(c) and (d), respectively, the orientations of which are in accordance with those proposed by Miyamoto {\it et al.} \cite{miyamoto}.
The C$_{20}$ monomers and dimers, which are individually optimized in advance \cite{comment-geometry}, are put between the electrodes.
Since many first-principles investigations using structureless jellium electrodes are in almost quantitative agreement with experiments \cite{Lang, Kobayashi, Okamoto, Tsukamoto}, we substitute jellium electrodes for crystal ones.
The distance between the edge atoms of inserted molecules and the jellium electrode is set at 0.91 a.u.


We first determine the stable structures of C$_{20}$ monomers and dimers suspended between the Au electrodes.
For structural optimization, a conventional supercell is employed under the periodic boundary condition in all directions.
The size of the supercell is $L_x=L_y=21.6$ a.u. and $L_z=L_{mol}+25$ a.u., where $L_x$ and $L_y$ are the lateral lengths of the supercell in the $x$ and $y$ directions parallel to the electrode surfaces, $L_z$ is the length in the $z$ direction, and $L_{mol}$ is the length of the inserted molecules.
We take a cutoff energy of 110 Ry, which corresponds to a grid spacing of 0.30 a.u., and a higher cutoff energy of 987 Ry in the vicinity of nuclei with the augmentation of double-grid points \cite{tsdg}.
Structural optimizations of the molecules between the electrodes are further implemented until the remaining forces acting on atoms are smaller than 1.65 nN.
Consequently, all models become shorter along the $z$ axis \cite{shorter}.

Next, we explore the electronic conductance and channel transmissions of the C$_{20}$ molecules at the zero-bias limit with the Landauer formula  G=tr({\bf T}$^\dagger${\bf T})G$_0$, where {\bf T} is a transmission matrix.
The eigenchannels are investigated by diagonalizing the Hermitian matrix ({\bf T}$^\dagger${\bf T}) \cite{Kobayashi}.
Here, we impose the nonperiodic boundary condition in the $z$ direction and employed semi-infinite Au jellium electrodes.
A cutoff energy (a higher cutoff energy around nuclei) is set at 62 (555) Ry.
The results of the conductances and channel transmissions at the Fermi level are collected in Table \ref{tbl:conductance}.
Although an isolated C$_{20}$ molecule has a small gap between highest occupied molecular orbital and lowest unoccupied molecular orbital  \cite{Kashenock}, considerable electron conductions are observed in the C$_{20}$ molecules suspended between the electrodes.
The high conduction property is attributed to a significant amount of charge transfer from the electrodes to molecules \cite{comment-ChargeTransfer}, which is localized at the interface between the electrodes and molecules.
As be seen in Table \ref{tbl:conductance}, there are four channels that actually contribute to electronic conduction in both the monomer models: the monomer models exhibit the high conductance of 1.57 G$_{0}$ for the double-bonded one and 0.83 G$_{0}$ for the single-bonded one.
The difference in the conductance between the two monomer models is caused by the number of atoms facing the jellium electrodes.
In the case of the dimer models, the conductances of the double-bonded one and the single-bonded one are 0.18 G$_0$ and 0.17 G$_0$, respectively.
In spite of more atoms facing the electrodes, the double-bonded dimer has an approximately equal conductance to that of the single-bonded one, which cannot be expected from the above results of the monomer models.
The higher conductance of the single-bonded dimer is consistent with the previous calculation that the band gap of an infinite single-bond chain is significantly smaller than that of a double-bonded one \cite{miyamoto}.
As for the conductance dependency on the number of molecules between electrodes, Roland {\it et al.} \cite{Roland} demonstrated that the conductance at the Fermi level does not decrease even when the number of molecules increases.
This discrepancy between our results and theirs is mainly brought about by the difference of the molecular orientations; we adopted the models of the most stable geometry.

Figure \ref{fig:ChannelElectronDistributions} shows channel electron distributions at the Fermi level, where the cross sections at the center of the models along the $z$ direction are depicted.
Scatterings of incident electrons are found at the entrances and exits of the molecules rather than within the C$_{20}$ cages.
Moreover, in the case of dimers, electrons are clearly reflected at the C$_{20}$-C$_{20}$ junctions.
Channel current distributions at the Fermi level are illustrated in Fig.\ref{fig:CurrentDensityDistributions}, in which electrons are observed to conduct along the C-C bonds of the C$_{20}$ cages rather than inside the cages.
We found a more striking feature that local loop currents are induced around the outermost carbon atoms.


In summary, we have investigated electron conduction properties of C$_{20}$ molecules connected to semi-infinite Au jellium electrodes using the overbridging boundary-matching method, which allows us to carry out first-principles transport calculations.
Our results indicate that conductances greatly depend on the number of C$_{20}$ molecules between the electrodes: the conductances are around 1 G$_{0}$ for the monomer models, and extremely small for the dimer models.
We find that the electron currents follow the C-C bonds, and do not cross the C$_{20}$ cages.
Incident electrons are predominantly scattered at both the C$_{20}$-electrode interfaces and the C$_{20}$-C$_{20}$ junctions while they are hardly reflected within the C$_{20}$ molecules.
In addition, the channel currents are observed to form local loops circulating around the outermost carbon atoms.
The present analysis is applicable to various kinds of nanostructures, and will enable the elucidation of their conduction properties.

This research was partially supported by the Ministry of Education, Culture, Sports, Science and Technology, Grant-in-Aid for Young Scientists (B), 14750022, 2002. The numerical calculation was carried out by the computer facilities at the Institute for Solid State Physics at the University of Tokyo, and Okazaki National Institute.


\newpage

\begin{figure*}[htbp]
\begin{center}
\caption{Schematic descriptions of the scattering region of C$_{20}$ molecules suspended between Au jellium electrodes: (a) double-bonded monomer, (b) single-bonded monomer, (c) double-bonded dimer, and (d) single-bonded dimer models.}
\label{fig:models}
\end{center}
\end{figure*}

\begin{figure*}[htbp]
\begin{center}
\caption{Channel electron distributions at the Fermi level: (a) double-bonded monomer, (b) single-bonded monomer, (c) double-bonded dimer, and (d) single-bonded dimer models. The planes shown are perpendicular to the electrode surfaces and contain the atoms facing electrodes. The states incident from the left electrodes are depicted. The circles, lines, and broken lines represent carbon atoms, C-C bonds, and the edges of the jellium electrodes, respectively.}
\label{fig:ChannelElectronDistributions}
\end{center}
\end{figure*}

\begin{figure*}[htbp]
\begin{center}
\caption{Channel Current distributions at the Fermi level: (a) double-bonded monomer, (b) single-bonded monomer, (c) double-bonded dimer, and (d) single-bonded dimer models. The planes shown are the same as in Fig.\ref{fig:ChannelElectronDistributions}. The circles, lines, and broken lines represent carbon atoms, C-C bonds, and the edges of the jellium electrodes, respectively.}
\label{fig:CurrentDensityDistributions}
\end{center}
\end{figure*}

\begin{table}[thbp]
\begin{center}
\caption{Conductances and channel transmissions at the Fermi level: (a) double-bonded monomer, (b) single-bonded monomer, (c) double-bonded dimer, and (d) single-bonded dimer models.}
\label{tbl:conductance}
\begin{tabular}{c|c|cccc} \hline\hline
\makebox[9mm]  & \makebox[30mm] & \multicolumn{4}{c}{Channel Transimissions} \\
                & \raisebox{2ex}{Conductance(G$_{0}$)} & \makebox[10mm]{1} &
  \makebox[10mm]{2}   &  \makebox[10mm]{3}   &  \makebox[10mm]{4}   \\ \hline
(a)             & 1.57                                   & 0.47 & 0.47 & 0.35 & 0.20 \\
(b)             & 0.83                                   & 0.36 & 0.18 & 0.17 & 0.06 \\
(c)             & 0.18                                   & 0.16 & 0.01 & 0.00 & 0.00 \\
(d)             & 0.17                                   & 0.07 & 0.06 & 0.03 & 0.01 \\ \hline\hline
\end{tabular}
\\
\end{center}
\end{table}


\begin{references}
\bibitem{comment-geometry}
We examined the most stable geometry of C$_{20}$ dimer among four types of dimers; a double-bonded, a single-bonded, CF1, and CF2 dimers. Then we concluded that the total energy of the double-bonded dimer is smaller than those of the single-bonded, CF1, and CF2 dimers, by 2.12, 0.88, and 2.60 eV, respectively. Here, CF1 and CF2 models are defined in Ref.\cite{Roland}.
\bibitem{shorter}
The decreases of length are 2.9\%, 13.5\%, 2.1\%, and 8.6\% for the double-bonded monomer, single-bonded monomer, double-bonded dimer, and single-bonded dimer models, respectively.
\bibitem{comment-ChargeTransfer}
The increases of electrons per C$_{20}$ molecule are 2.30\%, 3.15\%, 0.68\%, and 0.24\% for the double-bonded monomer, single-bonded monomer, double-bonded dimer, and single-bonded dimer models, respectively.
\end{references}
\end{document}